\documentclass[conference,10pt]{IEEEtran}
\IEEEoverridecommandlockouts
\usepackage[utf8]{inputenc}
\usepackage{newtxtext,newtxmath}
\usepackage[dvipsnames,svgnames,x11names,hyperref]{xcolor}
\usepackage[T1]{fontenc}
\usepackage{array}
\usepackage{url}
\usepackage{graphicx}
\usepackage{algorithmic}
\usepackage{microtype}
\usepackage{balance}
\usepackage[acronym,nohypertypes=acronym]{glossaries}

\ifCLASSOPTIONcompsoc
  \usepackage[caption=false, font=normalsize, labelfont=sf, textfont=sf]{subfig}
  \usepackage[nocompress]{cite}
\else
  \usepackage[caption=false, font=footnotesize]{subfig}
  \usepackage{cite}
\fi

\usepackage[capitalise]{cleveref}
\usepackage{balance}

\setacronymstyle{long-short}

\IEEEtriggercmd{\balance}
\IEEEtriggeratref{5}

\newcommand{\papertitle}{An Analysis of Amazon Echo's Network Behavior}

\newcommand{\cmd}[1]{\texttt{#1}}
\newcommand{\pathname}[1]{\texttt{#1}}

\newacronym{MITM}{MITM}{man-in-the-middle}
\newacronym{IoT}{IoT}{Internet of Things}
\newacronym{API}{API}{application programming interface}
\newacronym{LAN}{LAN}{local area network}
\newacronym{TLS}{TLS}{Transport Layer Security}
\newacronym{CA}{CA}{certificate authority}
\newacronym{CPU}{CPU}{central processing unit}
\newacronym{SD}{SD}{Secure Digital}
\newacronym{eMMC}{eMMC}{embedded MultiMediaCard}
\newacronym{PIN}{PIN}{personal identification number}
\newacronym{SIM}{SIM}{subscriber identity module}
\newacronym{AP}{AP}{access point}
\newacronym{WPA}{WPA}{Wi-Fi Protected Access}
\newacronym{HTTP}{HTTP}{Hypertext Transport Protocol}
\newacronym{DHCP}{DHCP}{Dynamic Host Configuration Protocol}
\newacronym{NAT}{NAT}{network address translator}
\newacronym{SIP}{SIP}{Session Initiation Protocol}
\newacronym{DNS}{DNS}{Domain Name System}
\newacronym{IP}{IP}{Internet Protocol}
\newacronym{HTTPS}{HTTPS}{Hypertext Transport Protocol Secure}
\newacronym{UART}{UART}{universal asynchronous receiver-transmitter}
\newacronym{USB}{USB}{Universal Serial Bus}
\newacronym{OOBE}{OOBE}{out-of-box experience}
\newacronym{P2P}{P2P}{peer-to-peer}
\newacronym{GO}{GO}{group owner}
\newacronym{SSID}{SSID}{service set identifier}
\newacronym{JSON}{JSON}{JavaScript Object Notation}
\newacronym{TCP}{TCP}{Transmission Control Protocol}
\newacronym{AES}{AES}{Advanced Encryption Standard}
\newacronym{CBC}{CBC}{cipher block chaining}
\newacronym{CMS}{CMS}{Cryptographic Message Syntax}
\newacronym{AVS}{AVS}{Alexa Voice Service}
\newacronym{LWA}{LWA}{Login With Amazon}
\newacronym{US}{US}{United States}
\newacronym{PSTN}{PSTN}{public switched telephone network}
\newacronym{UA}{UA}{user agent}
\newacronym{URI}{URI}{Uniform Resource Identifier}
\newacronym{GRUU}{GRUU}{Globally Routable UA URI}
\newacronym{STUN}{STUN}{Session Traversal Utilities for NAT}
\newacronym{ICE}{ICE}{Interactive Connectivity Establishment}
\newacronym{TURN}{TURN}{Traversal Using Relays Around NAT}
\newacronym{sRTP}{sRTP}{Secure Real-Time Transport Protocol}
\newacronym{SDES}{SDES}{Session Description Protocol Security Descriptions}
\newacronym{RTP}{RTP}{Real-Time Transport Protocol}
\newacronym{TTS}{TTS}{text-to-speech}
\newacronym{UI}{UI}{user interface}
\newacronym{URL}{URL}{Uniform Resource Locator}
\newacronym{SDP}{SDP}{Session Description Protocol}

\begin{document}
\title{\papertitle}

\author{
  \IEEEauthorblockN{%
    Jan Janak\IEEEauthorrefmark{1},
    Teresa Tseng\IEEEauthorrefmark{2},
    Aliza Isaacs\IEEEauthorrefmark{2},
    Henning Schulzrinne\IEEEauthorrefmark{1}%
  }
  \IEEEauthorblockA{%
    \IEEEauthorrefmark{1}Department of Computer Science, Columbia University, USA%
  }
  \IEEEauthorblockA{%
    \IEEEauthorrefmark{2}Barnard College, USA%
  }
  Email:
    janakj@cs.columbia.edu,
    \{ttt2148,ai2376\}@barnard.edu,
    hgs@cs.columbia.edu
}

\maketitle

\begin{abstract}
With over 20 million units sold since 2015, Amazon Echo, the Alexa-enabled smart speaker developed by Amazon, is probably one of the most widely deployed Internet of Things consumer devices. Despite the very large installed base, surprisingly little is known about the device's network behavior. We modify a first generation Echo device, decrypt its communication with Amazon cloud, and analyze the device pairing, Alexa Voice Service, and drop-in calling protocols. We also describe our methodology and the experimental setup. We find a minor shortcoming in the device pairing protocol and learn that drop-in calls are end-to-end encrypted and based on modern open standards. Overall, we find the Echo to be a well-designed device from the network communication perspective.
\end{abstract}

\glsresetall

\section{Introduction}\label{sec:introduction}

With over 20 million units sold since 2015~\cite{amazon-quaterly-results}, Amazon Echo, the Alexa-enabled smart speaker developed and sold by Amazon, is probably one of the most widely deployed \gls{IoT} consumer devices. The Echo found its way to many homes~\cite{smart-audio-report}, high school classrooms~\cite{alexa-higher-education}, and some hotels~\cite{alexa-hospitality}. Despite the very large installed base, surprisingly little is known about the device's network behavior. How secure is the Wi-Fi connection process? How secure is the connection to Amazon cloud? Are the calls made from the Echo encrypted? In this paper, we aim to shed some light on the device's encrypted network communication to answer the questions.

We obtained a first generation Amazon Echo and modified its firmware to make it vulnerable to \gls{MITM} attacks. We then connected the device to a \gls{MITM}[-capable] testbed where we could record, decrypt, and analyze the network traffic between the device, the companion smartphone application, and Amazon cloud.

We describe our hardware and firmware modifications, the methodology, and the design of the testbed. We then record and analyze the device paring protocol used between the Echo, the companion smartphone application, and Amazon cloud. Next, we decrypt and analyze the \gls{AVS} protocol, focusing on the differences from the publicly documented \gls{AVS} API available to third-party developers of Alexa-enabled products.

We also record, decrypt, and analyze the protocols used by the Echo's real-time drop-in communication feature (device calling and intercom). We find that this feature is based on modern standard protocols with custom authentication and authorization. We also find that  media streams are end-to-end encrypted and that the system is designed to keep the stream within the local (home) network where possible.

The primary contribution of this paper is an analysis and documentation of encrypted network communication of Amazon Echo. Specifically, we analyze: 1) the device pairing protocol (OOB), 2) the \gls{AVS} protocol, and 3) the Alexa drop-in calling signaling and media protocols. We also discuss some of the design tradeoffs and discovered limitations. Given the large installed base and potentially privacy-invasive nature of the device, we believe more information about the device's network behavior is needed.

The rest of the paper is organized as follows. In \cref{sec:related-work} we review literature and work related to Amazon Echo. We describe the device modifications, methodology, and the design of our experimental setup in \cref{sec:experimental-setup}. \cref{sec:network-behavior} analyses in detail the selected network protocols used by first generation Amazon Echo. In \cref{sec:discussion} we discuss some of the limitations and design tradeoffs discovered in the device pairing and drop-in calling protocols. We conclude and discuss the limitations of our approach and potential future work in \cref{sec:conclusion-future-work}.

\section{Related Work}\label{sec:related-work}

Clinton et al.~\cite{clinton:alexasurvey} performed a hardware analysis of first generation Amazon Echo. Firmware extraction from newer Echo models is discussed in~\cite{vasile2018breaking}. iFixit published a detailed teardown guide~\cite{echo-ifixit} for the device. We used some of these findings in our work: the description of debugging pads on base of the devices and the ability of the Echo to boot from an external \gls{SD} card. We used a firmware image obtained from~\cite{echohacking-wiki} and followed the steps outlined in~\cite{barnes2017:alexa} to gain access to the Echo's \gls{eMMC} filesystem.

A security analysis of the Echo is presented in~\cite{haack2017:alexa}. The authors performed a variety of attacks including a sound-based attack, \gls{PIN} brute-forcing attack, replay attack on network traffic, and an attempt to exploit Amazon Alexa's \gls{API}. While no major vulnerabilities were found, the authors discuss potential weaknesses in the 4-digit \gls{PIN} authentication method and in the wake-word detection algorithm. If successfully exploited, it may be possible to activate the Echo using a highly distorted sound sample which will not be recognized as a wake word by the user, but will be recognized as such by the Echo. This kind of attack could lead to privacy-invasive exploitation of the device.

The authors in~\cite{ford2018alexa} analyzed the network behavior of two Echo Dot devices over a 21-day period and created a network signature of the device from the captured traffic. The robustness of the signature will likely be limited due to a small number of devices and limited datasets.

The paper~\cite{chung2017digital} analyzes the client and cloud components of the Amazon Alexa ecosystem from a digital forensics perspective. The authors analyzed available artifacts and developed a proof-of-concept digital forensic tool with the ability to collect and analyze the artifacts. The tool uses official and unofficial \glspl{API} and collects artifacts from the cloud, Amazon Echo, and smartphone devices.

While there have been several studies~\cite{weinstein-echo,apthorpe2017smart} focusing on the Echo's hardware security and network behavior, little is known about the device's encrypted traffic. To the best of our knowledge, our study is the first attempt to document and analyze Echo's encrypted network communication.

\section{Methodology and Experimental Setup}\label{sec:experimental-setup}

\begin{figure}
    \centering
    \includegraphics[width=\linewidth]{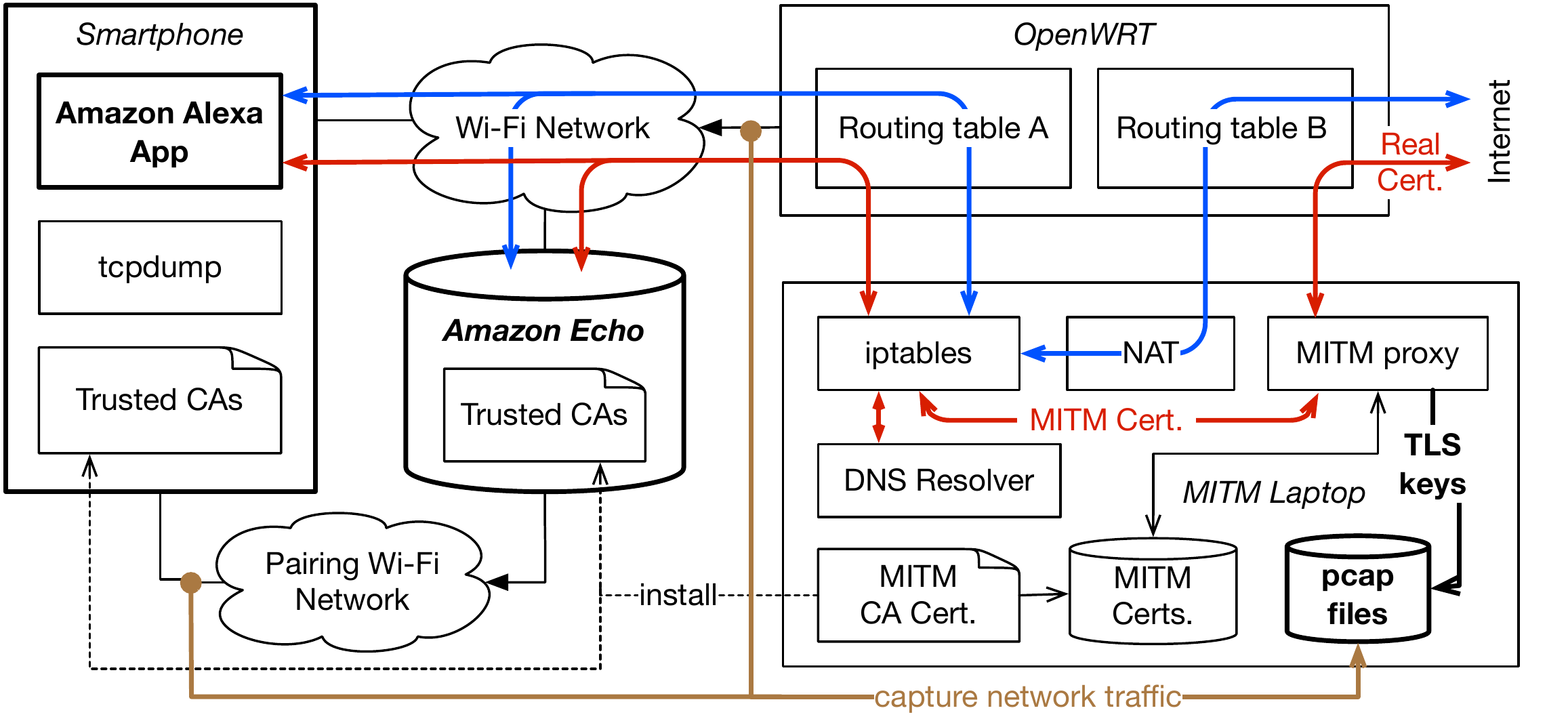}
    \caption{Experimental setup with Amazon Echo, Android smartphone with Alexa app, and a laptop with Mitmproxy. The Android device and the Echo have been modified to accept Mitmproxy's certificates.}\label{fig:setup}
\end{figure}

\cref{fig:setup} shows the architecture of our setup to capture and decrypt network traffic between Amazon cloud, Echo device, and Alexa Android application. We used a first generation Amazon Echo with software version 618622220. The Alexa application was running on a rooted Huawei Honor 6X phone with Android 7.0. Our Wi-Fi \gls{AP} was a Linksys WRT1900AC with OpenWRT 17.01.4. Mitmproxy 4.0 (\url{https://mitmproxy.org}) was running on a Lenovo X60 laptop with Ubuntu Linux 18.04. We used tcpdump on the rooted smartphone to capture pairing traffic between the Echo and the Alexa app. We also set Mitmproxy's \gls{CA} as trusted on the Android device.

To access encrypted network traffic between the Echo, Alexa app and Amazon cloud, we diverted all outgoing network traffic from the two devices to the laptop with iptables on the OpenWRT router. The laptop also served as a transparent \gls{NAT} and router for all the network traffic we did not want to decrypt. To decrypt \gls{TLS} traffic, we configured iptables on the laptop to forward such traffic to Mitmproxy. This was only applied to \gls{TLS} connections between the Echo or Alexa app and Amazon cloud. Mitmproxy performed a \gls{MITM} attack on the connections, replacing Amazon server certificates with locally generated ones. We used Wireshark to decrypt and analyze all captured network traffic.

\begin{figure}
    \centering
    \includegraphics[width=0.8\linewidth]{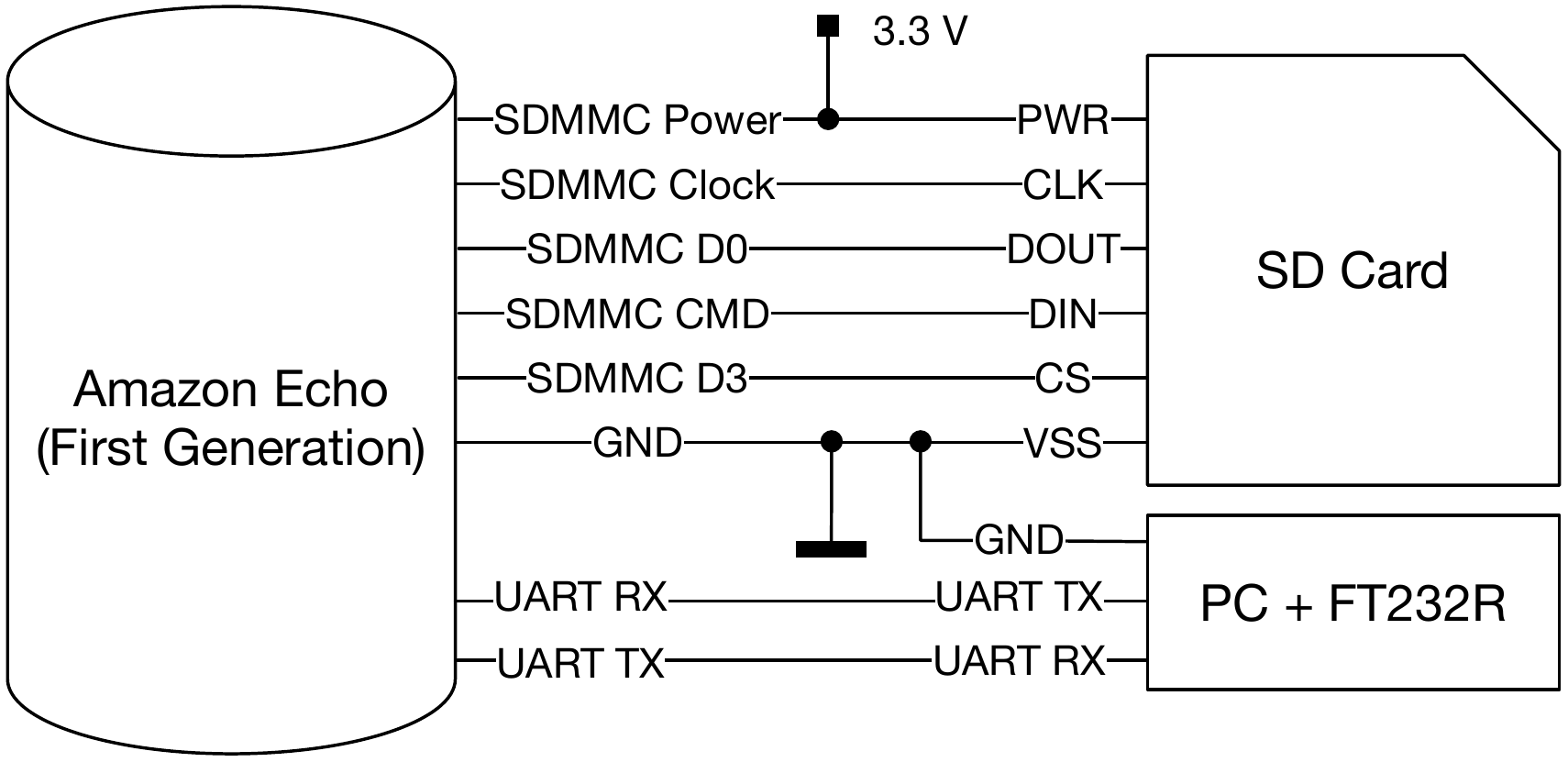}
    \caption{A modified Amazon Echo with an external SD card and a UART-USB interface attached to the device's debugging pads.}\label{fig:echo-sd-card}
\end{figure}

We modified the Echo device to make it accept Mitmproxy server certificates. All our hardware modifications are based on the work done by the authors in~\cite{barnes2017:alexa,echohacking-wiki}. We exposed the debugging pads at the bottom of the device and connected a UART-USB converter to gain access to the serial console used by the boot loader. We then connected an external \gls{SD} card to the remaining pads. The entire hardware modification is shown in \cref{fig:echo-sd-card}.

We then loaded the firmware image from~\cite{echohacking-wiki} on the external \gls{SD} card and rebooted the device. The custom firmware had a command line interface enabled, which allowed us to change the boot procedure. We followed the steps from~\cite{barnes2017:alexa} to obtain a root shell. In the root shell, we appended the Mitmproxy \gls{CA} certificate to file /etc/ssl/certs/ca-certificates.crt. After another reboot, the Echo would treat \gls{TLS} connections modified by Mitmproxy as legitimate, allowing us to decrypt the communication.

\section{Network Behavior}\label{sec:network-behavior}

In this section, we document and analyze three network protocols used by Amazon Echo: the device pairing protocol (OOBE), the \gls{AVS} protocol, and the drop-in calling protocol. All three protocols are partially or fully encrypted with \gls{TLS} on the network.

\subsection{Device Pairing Protocol (OOBE)}\label{sec:device-pairing}

The \gls{OOBE} is a pairing protocol used to: a) provision a network credential (Wi-Fi network name and password) into the Echo device, and b) to associate the device with an Amazon account. This protocol is executed between the Echo, a pairing client in the form of Amazon Alexa smartphone app or a web application, and backend services in Amazon cloud. Pairing must be performed after the device has been reset to factory defaults, or when the Wi-Fi network becomes unusable, e.g., due to a Wi-Fi password change. Pairing can also be manually activated by pressing and holding down the ``dot'' button on the Echo.

\begin{figure}
  \centering
  \includegraphics[width=0.8\linewidth]{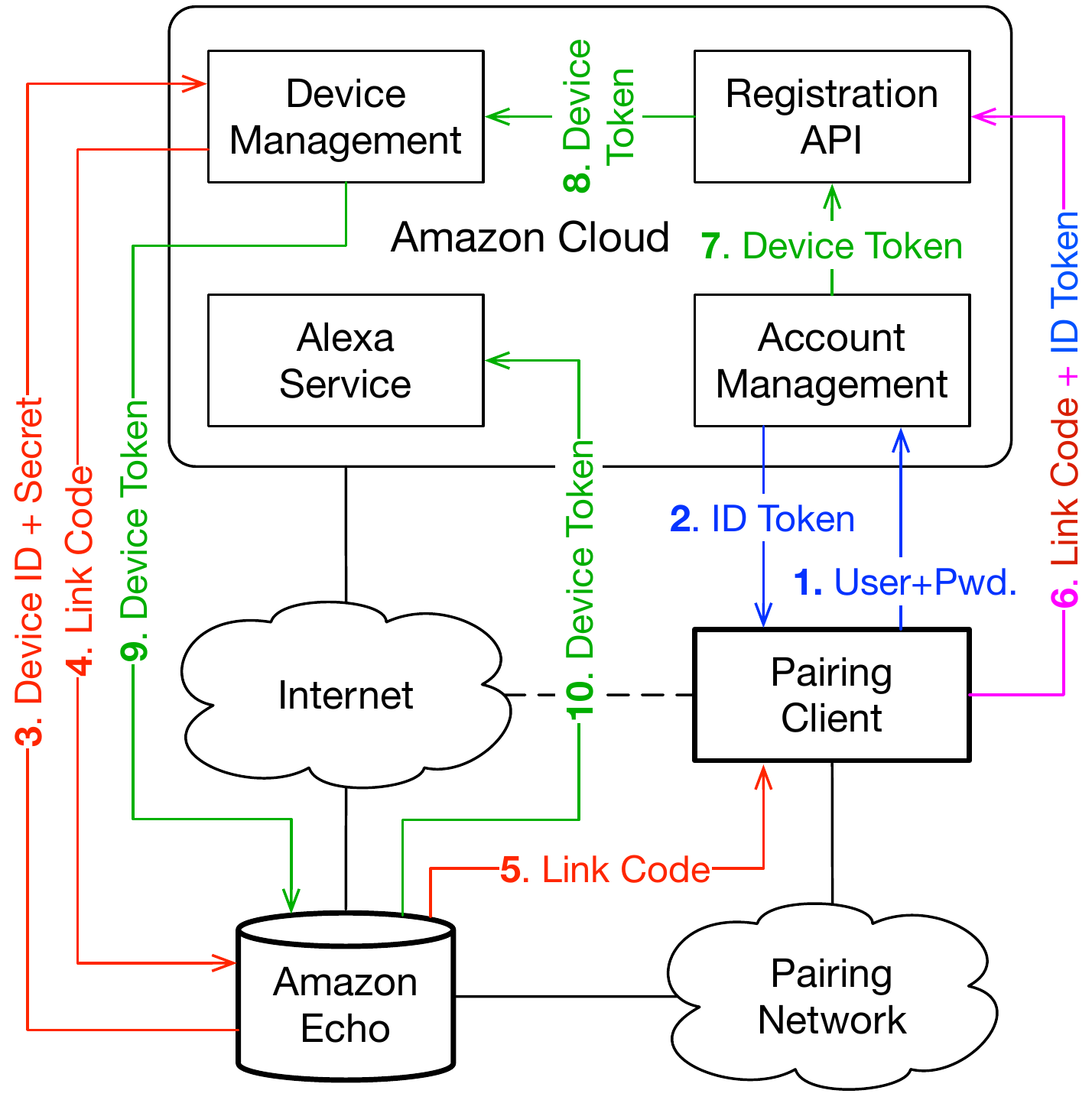}
  \caption{Data flow diagram of the device pairing protocol (OOBE).}\label{fig:oobe-data-flow}
\end{figure}

\begin{figure}
  \centering
  \includegraphics[width=\linewidth]{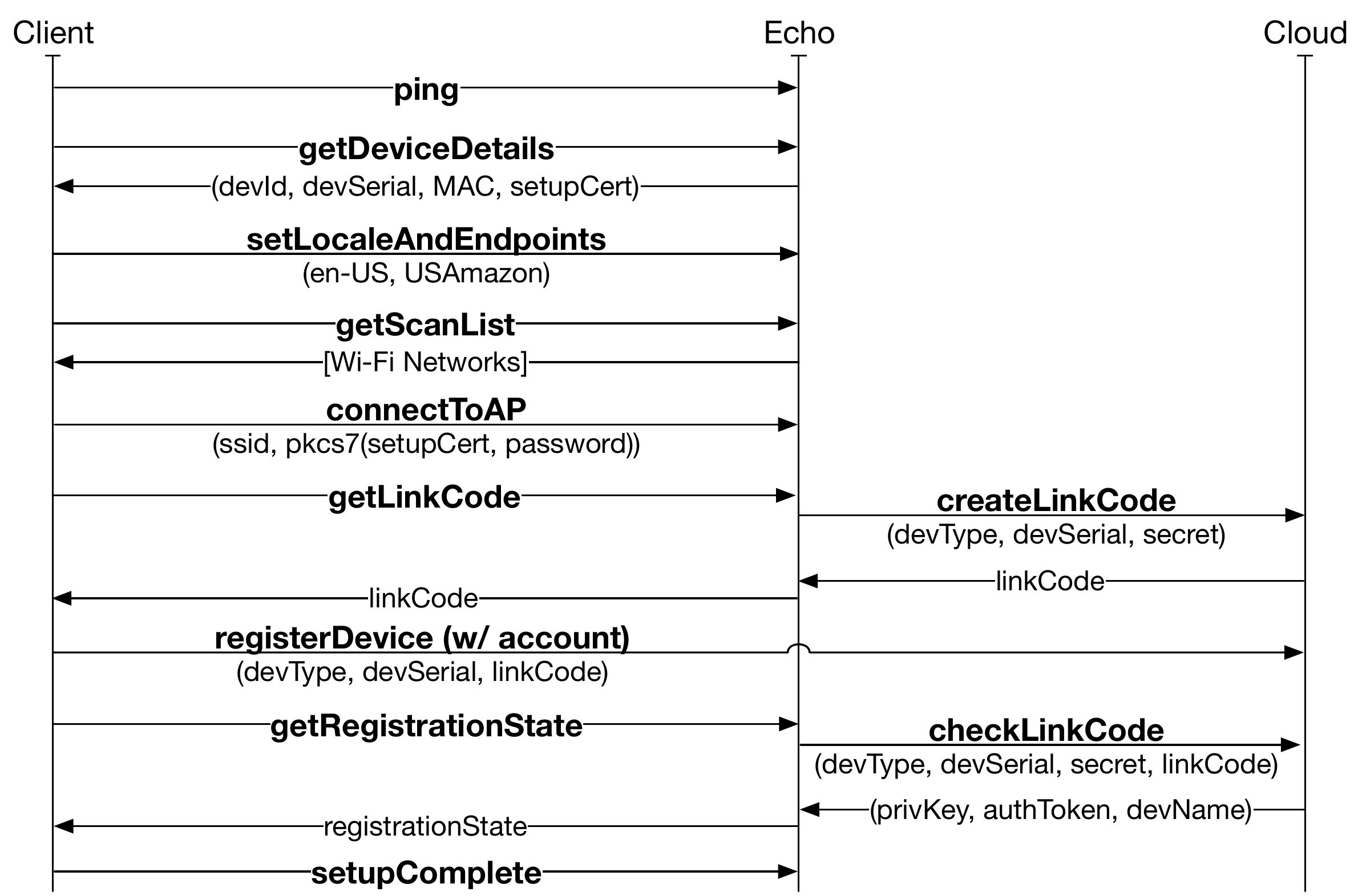}
  \caption{Call flow diagram of the device pairing protocol (OOBE).}\label{fig:oobe-call-flow}
\end{figure}

The pairing exchange takes place over an open temporary Wi-Fi network created by the Echo. The device creates a Wi-Fi \gls{P2P} group and configures itself as the \gls{GO}. The \gls{SSID} of the temporary network is ``Amazon-XYZ'' where XYZ represents three digits from the device's serial number. The Echo can create a temporary Wi-Fi network regardless of the state of its main Wi-Fi interface, i.e., even when connected as client to another network on a different channel.

On the temporary network, the Echo configures itself with a fixed IP address, starts DHCP and DNS servers, and configures iptables to redirect all DNS requests from connected clients to its internal DNS server. The internal DNS server has a few well-known Amazon hostnames and IP addresses hard-wired, presumably to be able to resolve those even when the device itself is not connected to the internet. If connected to the internet during pairing, the Echo forwards (\gls{NAT}[-ed]) traffic from the pairing client to the internet via its main Wi-Fi interface. This is used by the pairing client to associate the device with the user's Amazon account. The Echo audibly announces only the first connected pairing client with no identification. Multiple pairing clients can be connected simultaneously.

\cref{fig:oobe-data-flow} shows the data exchanged between the entities involved in the pairing process. \cref{fig:oobe-call-flow} shows the interaction between the pairing client, Echo, and Amazon cloud. The Echo provides a pairing API on TCP ports 8080 and 443. A HTTP server on port 8080 accepts JSON-serialized Thrift~\cite{thrift} messages for the path \pathname{/OOBE}. Port 443 is a built-in TCP proxy that forwards HTTPS connections from the pairing client to Amazon cloud.

The pairing client periodically invokes the \cmd{ping} API method to determine if it is connected to the correct network on an Echo in pairing mode. Since different Echo models may use different static IPs, the client sends the request to several pre-configured IPs simultaneously. Pairing is initiated with the first device that correctly acknowledges the ping request. The ping method serves as a rudimentary service discovery mechanism. Unlike, e.g., ZeroConf, the ping method can be used by browser applications such as the pairing application available at \url{https://alexa.amazon.com}.

Upon discovering an Echo in pairing mode (indicated by spinning orange light), the client invokes the \cmd{getDeviceDetails} method. The method returns basic device information: device model, serial number, Wi-Fi hardware address, supported languages, and a pairing X.509 public key certificate (used later). The client then selects the language and configures Amazon cloud endpoints corresponding to the device's geographic location.

Next, the client determines whether the Echo has any Wi-Fi credentials configured already and obtains the list of scanned Wi-Fi networks. In \cref{fig:oobe-call-flow}, these two steps are represented by the \cmd{getScanList} request. The client prompts the user to select a Wi-Fi network and invokes \cmd{connectToAP} to connect the Echo to the selected network.

The pairing client encrypts the Wi-Fi credential with AES-256 in CBC mode using a random secret key. The key is then encrypted with the public key from the X.509 certificate returned by \cmd{getDeviceDetails}. The encrypted key and Wi-Fi credential are then transmitted in the \gls{CMS} format~\cite{rfc5652}. The X.509 certificate provided by the Echo is self-signed and this method is thus vulnerable to active \gls{MITM} attacks. The encrypted message can be decrypted with the openssl command:

\begin{small}
\begin{verbatim}
openssl smime -decrypt -in <data> -inform pem \
  -inkey /var/local/oobe-web-setup.cert
\end{verbatim}
\end{small}
File \pathname{/var/local/oobe-web-setup.cert} can be obtained from Echo's firmware image.

The pairing client then obtains a \emph{link code} from Amazon cloud via the Echo (steps 3--5 in \cref{fig:oobe-data-flow}). The Echo first submits its device type, serial number, and a secret string to the \cmd{createLinkCode} API in Amazon cloud. The returned link code (step 4 in \cref{fig:oobe-data-flow}) consists of five alphanumeric characters and represents the device during registration with an Amazon user account. The format suggests that the code may have been designed for scenarios where the user is expected to transmit the string manually from the device to Amazon via the web interface. That API appears to be part of a device rendezvous service, most likely connected to an inventory database that keeps track of the serial numbers and secrets for all manufactured devices. The secret string appears to be set on each Echo during the manufacturing process.

Having received the link code, the pairing client registers the device with the user's Amazon account (step 6 in \cref{fig:oobe-data-flow}). Since the client is connected to the Echo's pairing Wi-Fi network, it uses the device as a proxy to reach Amazon cloud in order to register the device. The client's request is end-to-end encrypted with TLS and authenticated with a HTTP cookie that authenticates the user (obtained in steps 1--2 in \cref{fig:oobe-data-flow}). The client submits the device type, serial number, and the link code identifying the device to be registered with the authenticated user's account. Amazon cloud verifies the link code and associates the device with the user's account (steps 7--8 in \cref{fig:oobe-data-flow}). These last steps are entirely implemented by Amazon cloud and we were not able to obtain more information about them.

The client invokes \cmd{getRegistrationState} on the Echo to determine registration status. Using its secret, the Echo queries the rendezvous service for the state of the link code by periodically posting to the \cmd{checkLinkCode} API. Once registered, this API returns a private key, an authentication token (step 9 in \cref{fig:oobe-data-flow}), and the human-friendly name assigned to the device by the user. The private key and the authentication token are used by the Echo to access Amazon cloud on behalf of the user. The authentication token is included in every HTTP request to Amazon services (step 10 in \cref{fig:oobe-data-flow}) and appears to include data encrypted with a symmetric key wrapped with a public-private key known only to Amazon and an initialization vector.

The client terminates the pairing process by invoking the \cmd{setupComplete} method on the Echo, instructing the device to shut down the temporary Wi-Fi network.

\subsection{Alexa Voice Service (AVS)}\label{sec:alexa-voice-service}

The Alexa Voice Service (AVS) is an Amazon cloud service that provides speech recognition, natural language understanding, and text to speech capabilities to Amazon Echo devices. In this section, we briefly describe the proprietary AVS protocol used by our first generation Amazon Echo. \gls{AVS} is also available to third-party developers of Alexa-enabled products in the form of a public API. The public API is extensively documented online~\cite{amazon-alexa-developer}, is based on HTTP/2, and uses Amazon's \gls{LWA} authorization service~\cite{lwa}.

\begin{figure}
    \centering
    \includegraphics[width=0.8\linewidth]{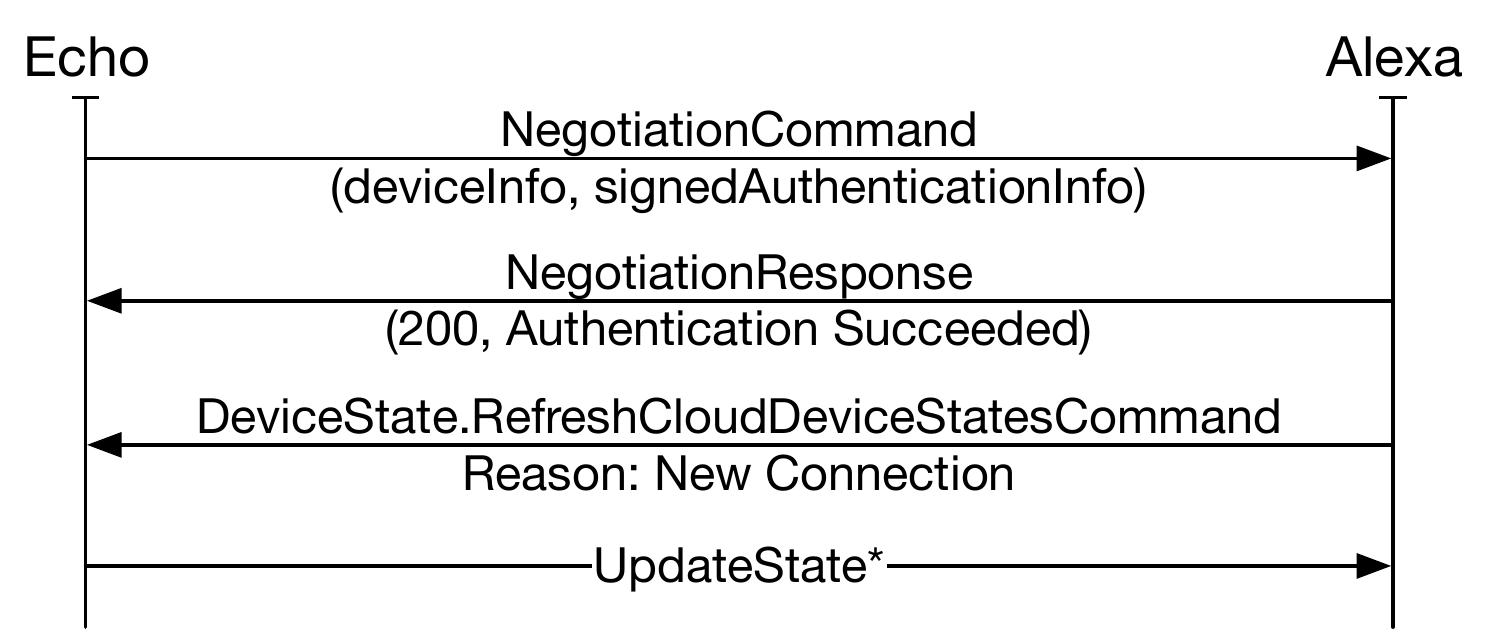}
    \caption{The setup phase of the persistent connection between the Echo and Alexa Voice Service (AVS).}\label{fig:avs-setup}
\end{figure}

Each Echo maintains a persistent long-term SPDY~\cite{spdy} connection to an Echo-specific \gls{AVS} endpoint. \cref{fig:avs-setup} shows the connection setup phase. The first command sent by the Echo is \cmd{NegotiationCommand} which authenticates the device. A portion of the command's JSON payload is cryptographically signed with the private key obtained by the device during pairing. The signed portion contains device type and serial number, an authentication token (also obtained during pairing), and a timestamp. The server then immediately notifies the Echo to refresh the state of various subsystems.

Further communication over the connection is similar to the public \gls{AVS} protocol described in~\cite{amazon-alexa-developer}, with the exception that the Echo provides interfaces not accessible via the public \gls{API} such as the SipClient interface described in \cref{sec:real-time-communication}. We omit the rest of the protocol due to space constraints and refer interested readers to the public AVS documentation.


\subsection{Alexa Drop-in Calling}\label{sec:real-time-communication}

Alexa drop-in calling can be used to place calls to other Alexa-enabled Echo devices, to the Alexa smartphone application, to selected U.S. phone numbers, or to Skype. In this section, we take a closer look at the protocols used by this feature. Uttering a phrase such as ``Alexa drop in on~\dots'', ``Alexa call~\dots'', or ``Alexa answer'' causes the device to establish a two-way call with another Alexa-enabled device, a phone number, or a Skype account. The system supports two call types: regular call and intercom.

Regular calls use the caller's address book, provided to the system by the Alexa smartphone app, to look up the callee. If the callee has Alexa-enabled devices, the call is routed to those devices. All devices indicate the incoming call simultaneously, but only one can answer the call. If the callee does not have any Alexa-enabled devices, the call is routed to a \gls{PSTN} or Skype gateway.

In intercom mode, the call is established between Alexa-enabled devices without the need for the callee to answer the call. The called device answers automatically, provided that the callee has granted a ``drop-in'' permission to the caller. The caller must specify a particular device to call, not a user or phone number. Often, this would be another Echo paired with the same Amazon account, e.g., an Echo in another room.

The drop-in calling feature is entirely based on open standards. Compatible devices run a \gls{SIP}~\cite{rfc3261} \gls{UA} based on PJSIP v2.7.1. Signaling is based on \gls{SIP} with Outbound~\cite{rfc5626}, Path~\cite{rfc3327}, and \gls{GRUU}~\cite{rfc5627} extensions. A media path for each call is negotiated with \gls{STUN}~\cite{rfc5389}, \gls{ICE}~\cite{rfc8445}, and \gls{TURN}~\cite{rfc5766} protocols. Audio is encoded with the Opus codec~\cite{rfc6716} and the stream is end-to-end encrypted with AES-256 using \gls{sRTP}~\cite{rfc3711} and \gls{SDES}~\cite{rfc4568} protocols.

The \gls{UA} is fully managed by the cloud-based Alexa service over its persistent (SPDY) control connection. Shortly after device start, the \gls{UA} sends a \cmd{ConfigureCommsRequest} provisioning request to Alexa which provides the \gls{UA} with \gls{SIP} registration configuration. The configuration includes the \gls{SIP} username, registrar domain, and a registration authorization credential. The \gls{UA} establishes and maintains a persistent \gls{TLS} connection to Amazon's \gls{SIP} service on non-standard port 443 (HTTPS). We assume the port has been chosen to facilitate firewall traversal. The \gls{UA} registers with the \gls{SIP} service using the configuration provided by Alexa. All devices associated with the same Amazon user account register the same \gls{SIP} URI. Each \gls{UA} is also assigned a device-specific \gls{SIP} URI. The account-specific \gls{SIP} URI is used to simultaneously reach all devices during regular calls. The device-specific \gls{SIP} URI is used to reach a particular device during intercom calls.

\begin{figure}
    \centering
    \includegraphics[width=1.0\linewidth]{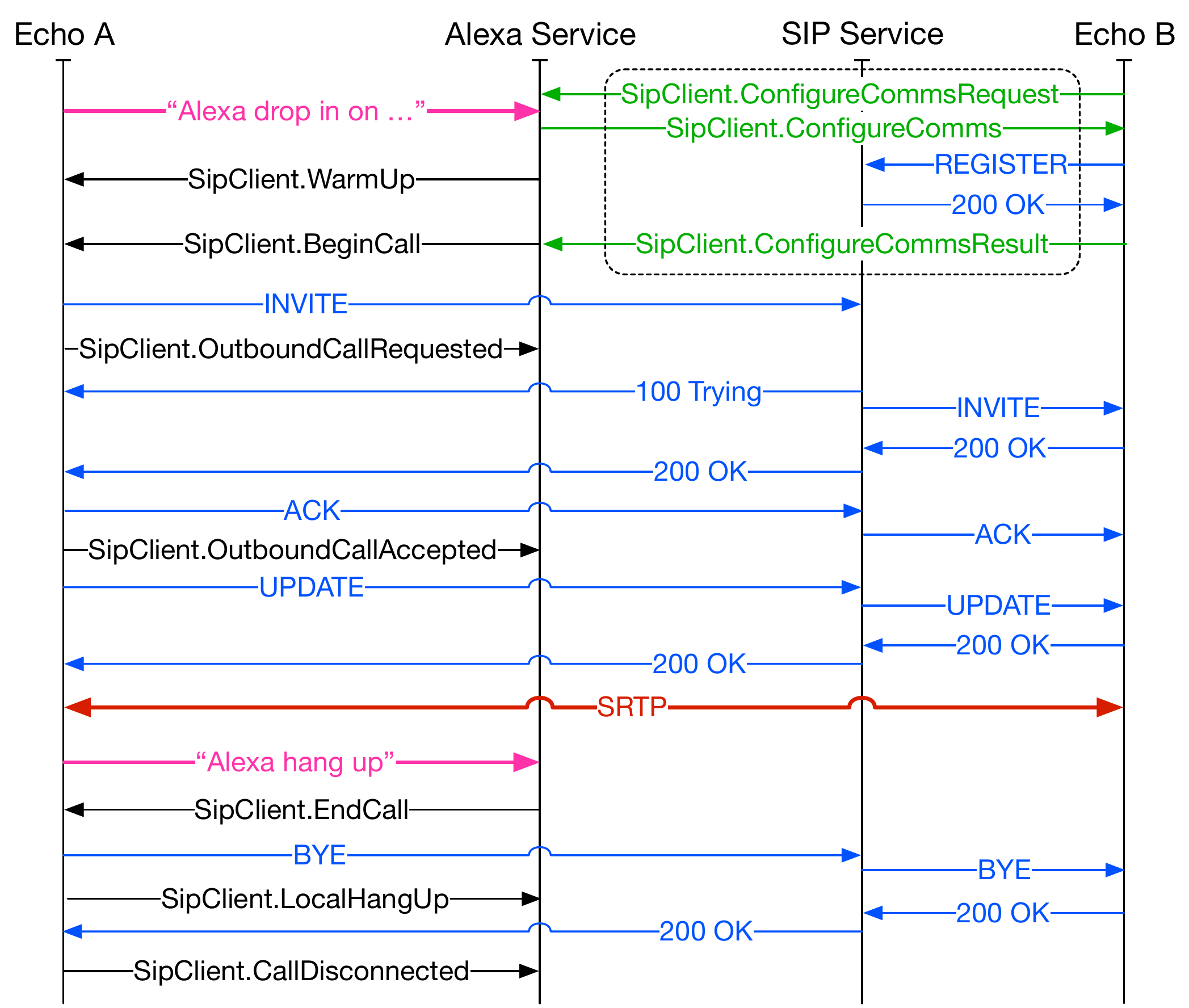}
    \caption{Intercom call flow diagram (utterances in purple, SIP in blue, Alexa protocol in black). Dashed rectangle represents startup registration signaling.}\label{fig:intercom-callflow}
\end{figure}

Having detected an utterance to establish a call, the Alexa service issues a \cmd{SipClient.WarmUp} command followed by a \cmd{SipClient.BeginCall} command to the \gls{UA}. The latter command includes JSON payload with detailed call-related configuration: the caller and callee \gls{SIP} URI, a call authorization token, available \gls{TURN} and \gls{STUN} endpoints, and various Alexa-specific attributes. \cref{fig:intercom-callflow} shows an intercom call flow diagram. The calling \gls{UA} sends a \gls{SIP} INVITE to the callee and notifies Alexa that the call is being established with a \cmd{SipClient.OutboundCallRequested} message. Once the call has been accepted, the \gls{UA} notifies Alexa again with a \cmd{SipClient.OutboundCallAccepted} message. Both notifications carry payload describing the state of the \gls{UA}.

Both media streams between two Echo devices are end-to-end encrypted. As is common in \gls{SIP}[-based] systems, the encryption key is derived from a master key exchanged in \gls{SIP} signaling. Since all \gls{SIP} signaling takes place over authenticated TLS connections, the key is secure against eavesdropping attacks. However, Amazon's \gls{SIP} service has access to the master key (by design) and can thus decrypt the media streams.

\begin{figure}
    \centering
    \includegraphics[width=0.7\linewidth]{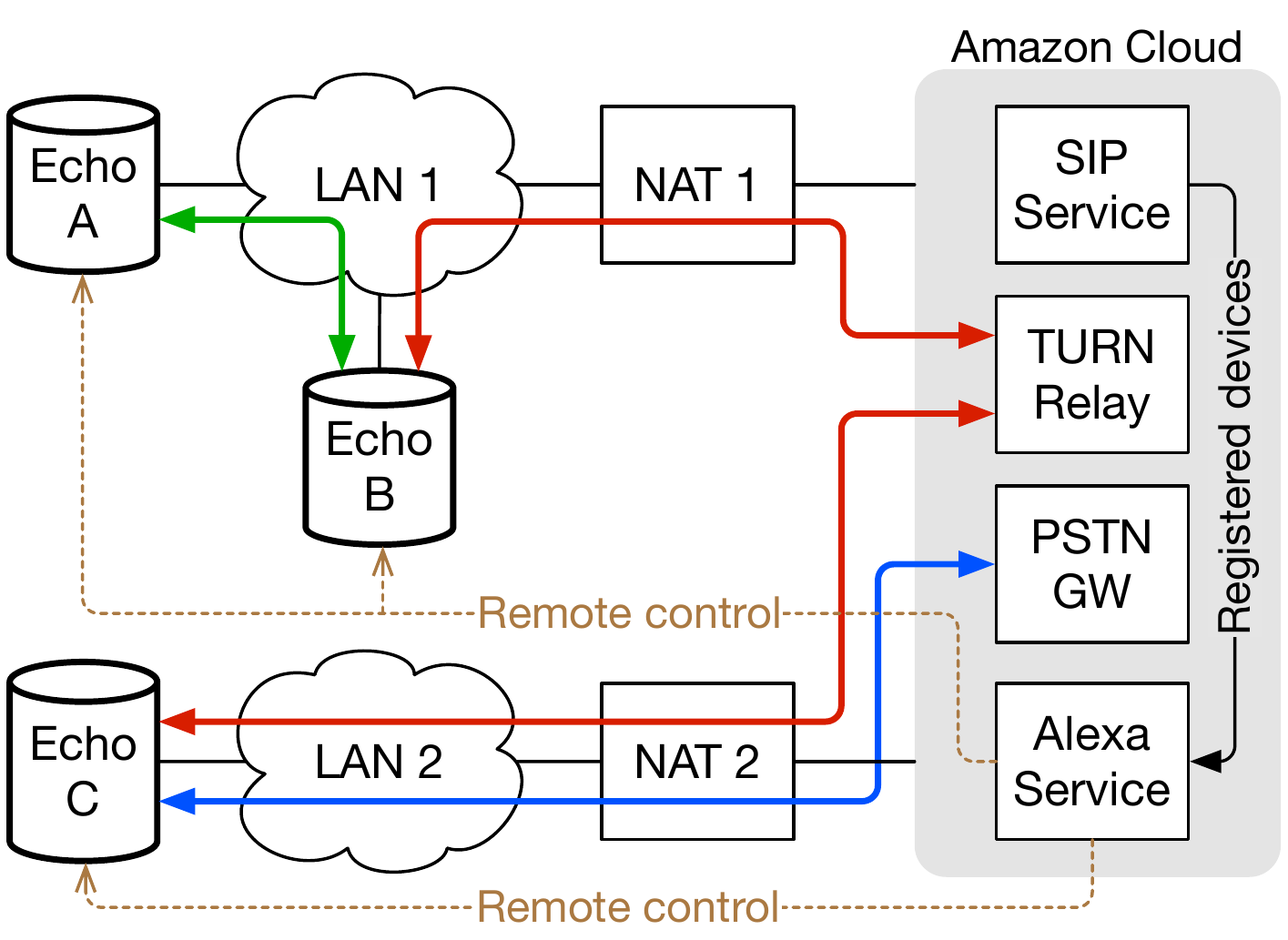}
    \caption{Architecture of the calling subsystem. SIP, TURN, and PSTN services are provided by Amazon cloud. Calls between devices in the same LAN stay local (green). Others are relayed (red) or terminated in the cloud (blue).}\label{fig:calling-architecture}
\end{figure}

Whether media passes through Amazon cloud depends on network configuration. \cref{fig:calling-architecture} shows an architecture diagram with three possible cases. A call established between two devices in the same LAN, as is common for intercom calls between two devices in the same home, remains in the LAN. A call placed to a device behind another home router will likely have media streams forced through a \gls{TURN} relay in Amazon cloud. This is common when calling another Amazon user. Calls to \gls{PSTN} or Skype are always terminated at a cloud-based gateway.

Uttering ``Alexa hang up'' terminates the call. Alexa sends the request \cmd{SipClient.EndCall} to the \gls{UA} which sends a \gls{SIP} BYE. After the call ended, the \gls{UA} notifies Alexa with a \cmd{SipClient.CallDisconnected} message.

The drop-in calling feature uses a custom \gls{SIP} authentication method tied to the Alexa service. Each \gls{SIP} request carries a proprietary \texttt{X-authtoken} header with a short-lived authorization token generated by Alexa. A portion of the token is signed by a private key tied to the Amazon user account. Only the Alexa service has access to the key. Since the authorization token is cryptographically bound to both calling and called \gls{SIP} URIs, the \gls{UA} cannot initiate a call unless authorized by Alexa. Each call needs a new authorization token.

\section{Discussion}\label{sec:discussion}

The modification shown in~\cref{fig:echo-sd-card} is only effective with the first generation Echo hardware. Consequently, the \gls{MITM} attack described in \cref{sec:experimental-setup} is harder to perform on newer Echo models where the firmware applies additional security measures. Thus, installing a custom \gls{CA} certificate on the device is harder. However, the network protocols analyzed in this paper are compatible with newer Echo models and other parts of the ecosystem.

Neither the temporary Wi-Fi network nor the pairing protocol are encrypted and are vulnerable to eavesdropping. An attacker who could observe the pairing exchange could obtain the link code and might be able to associate a previously de-registered device with different Amazon account. A newly purchased Echo comes pre-registered to the Amazon user account used to purchase the device. This mechanism effectively prevents hijacking an Echo using the pairing protocol, as long as the device is already registered in Amazon cloud.

One of the supported pairing clients is a web application implemented in JavaScript. In order for the client to be able to communicate with the Echo device via unencrypted HTTP, the client itself must be served via unencrypted HTTP. This is necessary to work around the security limitations imposed by modern web browsers. The web application is first loaded via HTTPS and after the user has logged in, the browser is redirected to a HTTP URL which downloads the JavaScript pairing client. This design leaves the JavaScript pairing client vulnerable to code injection by remote attackers.

The custom authentication mechanism used in drop-in call signaling could use more scrutiny, e.g., to see whether it might be vulnerable to well-known attacks on \gls{SIP}[-based] systems such as \gls{SIP} header substitution, downgrade, or media encryption downgrade. Since the intercom feature in Amazon Echo answers incoming calls automatically, potential vulnerabilities in the design of Echo's \gls{SIP} infrastructure could turn an Echo device into a remotely controllable microphone.

\section{Conclusion and Future Work}\label{sec:conclusion-future-work}

Despite the very large installed base, not much is known about Amazon Echo's network behavior. In this paper, we analyzed and documented the device pairing protocol (OOBE), the Amazon Voice Service (AVS) protocol, and the Alexa drop-in calling protocol used by a first generation Amazon Echo. We modified the firmware to make the device vulnerable to \gls{MITM} attacks. We then mounted a \gls{MITM} attack against the device and decrypted the communication between the device, pairing smartphone application, and Amazon cloud. We also described the approach and the experimental setup in detail.

We found limitations in the device pairing protocol (OOBE) which under certain circumstances could be used to associate a de-registered Echo with a different Amazon account. This vulnerability is effectively mitigated by the pre-existing association of a new device to the purchasing Amazon account. Intercom calls are authorized with one-time authorization tokens issued by the Alexa service. Both signaling and media are end-to-end encrypted and use modern industry standard protocols. Overall, we find the first generation Amazon Echo to be a well-designed device from the network communication perspective.

Our experiments were limited to the first generation Amazon Echo. Later models make the \gls{MITM} approach considerably more difficult. This limitation does not change the analysis of network communication which is the primary contribution of this paper. The analyzed protocols are compatible and interoperable with more recent Amazon Echo models.

Analyzing the network communication employed by more recent Echo models, other smart speaker brands, or analyzing the communication with other \gls{IoT} devices on the same LAN is left for future work.

\bibliographystyle{IEEEtran}
\bibliography{bibs/references,bibs/ietf/rfc}
\end{document}